\documentclass[twocolumn,showpacs,showkeys,preprintnumbers]{revtex4-1}
\usepackage{amssymb}

\usepackage{amsfonts}

\usepackage{latexsym}

\usepackage{dcolumn}

\usepackage{amsmath}

\usepackage{graphicx}

\usepackage{psfrag,pstricks}

\begin{document}

\title{ Steady state entanglement, cooling, and tristability in a nonlinear optomechanical cavity}

\author{S.Shahidani $^{1}$ }
\email{sareh.shahidani@gmail.com}

\author{M. H. Naderi$^{1,2}$}
\author{ M. Soltanolkotabi$^{1,2}$}
\author{Sh. Barzanjeh$^{1,3}$}
\affiliation{$^{1}$ Department of Physics, Faculty of Science, University of Isfahan, Hezar Jerib, 81746-73441, Isfahan, Iran\\
$^{2}$ Quantum Optics Group, Department of Physics, Faculty of Science, University of Isfahan, Hezar Jerib, 81746-73441, Isfahan, Iran\\
$^{3}$School of Science and Technology, Physics Division,
Universita di Camerino, I-62032 Camerino (MC), Italy}
\date{\today}

\begin{abstract}
The interaction of a single-mode field with both a weak Kerr medium and a parametric nonlinearity in an intrinsically nonlinear optomechanical system is studied. The nonlinearities due to the optomechanical coupling and Kerr-down conversion  lead to the  bistability and tristability   in the mean intracavity photon number.
Also, our  work   demonstrates that the lower bound of the resolved sideband regime and the minimum attainable phonon number can be less than that of a bare cavity by controlling the parametric nonlinearity and the phase of  the driving field. 
   Moreover, we find that in the system under consideration  the degree of entanglement between the mechanical and optical modes is dependent on the  two stability parameters of the system. 
For  both cooling and entanglement, while parametric nonlinearity increases the optomechanical coupling , the weak Kerr nonlinearity is very useful  for  extending  the domain of the stability  region to  the desired range in which the minimum effective temperature and maximal  entanglement are  attainable. Also,  as shown in this paper,  the present scheme allows to have significant entanglement in the tristable regime for the lower and middle branches which makes the current scheme distinct from the bare optomechanical system. 
\end{abstract} 


\maketitle

\section{Introduction}

Considerable interest has recently been focused on the  optomechanical system as an  excellent candidate for studying the transition of a macroscopic degree of freedom from the classical to the quantum regime. This   system also provides   novel routes  for practical  applications such as  detection and interferometry of gravitational waves\cite{braginsky}  and  quantum limited displacement sensing \cite{rugar}. 
The standard and simplest setup of this  system is a Fabry-Perot cavity in which one of the mirrors is much lighter than the other, so that it can move under the effect of the radiation pressure force. 

State-of-the-art  technology   allows experimental demonstration of cooling of the vibrational mode of the  mechanical  oscillator to its ground state\cite{connel, teufel,chan} and  strong optomechanical coupling between the vibrational mode of the  mechanical
oscillator and   cavity field\cite{arcizet, gigan, metzger, groblacher}. This coupling is intrinsically nonlinear since the length of the cavity depends upon the intensity of the field in analogous way to the optical length of a Kerr material\cite{nori}. Therefore it    enables pondermotive squeezing of the field\cite{fabre}, photon blockade \cite{rabl}, generation of nonclassical states of the mechanical and optical mode \cite{nunnenkamp}, optical bistability\cite{dorsel} and  phonon-photon entanglement in the bistable regime \cite{ghobadi}. 
  Besides this intrinsic nonlinearity,  the presence of an optical parametric amplifier (OPA) \cite{huang, xuereb} or  the optical Kerr medium \cite{kumar} inside the cavity   has opened up a new domain for combining nonlinear optics and optomechanics towards the enhancement of quantum effects. It has been predicted \cite{huang} that the presence of  an OPA in a single-mode Fabry-Perot cavity causes a strong coupling between the oscillating mirror and the cavity mode resulting from increasing  the intracavity photon number. Also, when the optomechanical cavity contains an optical Kerr medium with strong $ \chi^{(3)} $ nonlinearity,  the photon-photon repulsion and the reduction of the cavity photon fluctuations provide a feasible route towards  controlling the dynamics of the
micromirror\cite{kumar}.

On the other hand,  the interaction of a single-mode field with both a Kerr medium and a parametric nonlinearity is a well-known quantum optical model which has been proposed for the generation of  nonclassical  states of the cavity field \cite{wielinga, leonski}.

Here,  we consider  the interaction of a single-mode field with both a weak Kerr medium and a parametric nonlinearity in an intrinsically nonlinear optomechanical system. In particular, we investigate the multistability , intensity, back-action cooling and stationary optomechanical  entanglement. 
 It turns out that the mean intracavity photon number,  in addition to bistability, 
exhibits tristability  for a certain range of the parameters which can be controlled by the intrinsic and external  nonlinearities in the system. 
Then,  we investigate the effect of Kerr-down conversion nonlinearity on the back-action ground state cooling of the mirror based on the covariance matrix and identify the modified optimal regime for  cooling. We will show that the lower bound of the resolved sideband regime and the minimum attainable phonon number can be less than that of a bare cavity by controlling the parametric nonlinearity and the phase of  the  field driving the OPA.  Also, the weak Kerr nonlinearity can be used to  extend the domain of the stability  to the desired range of the effective detuning in which the effective temperature of the system  minimizes. 
Then we show that  for a fixed effective detuning, in spite of  the  bare cavity,  one of  the stability parameters    (the counterpart of the bistability parameter) is a nonlinear function of the input power,  allows to approach  significant entanglement  simultaneously with the ground state cooling  of the mirror. In the last part  of the paper we shall focus on the generation of stationary entanglement in the presence of the nonlinearity and show  that not only  the entanglement is not a monotonic function of the optomechanical coupling strength, but also the two stability parameters of the system are the   key parameters  for maximizing  the degree of  entanglement. 
Based on these results we show that in the tristable regime, for the first and second branches  the degree of  entanglement is maximum at the end of the branches, while for the third branch the phonon-photon entanglement is null.


\section{The Physical Model}\label{sec1}
\begin{figure}[ht]
\centering
\includegraphics[width=3.2 in]{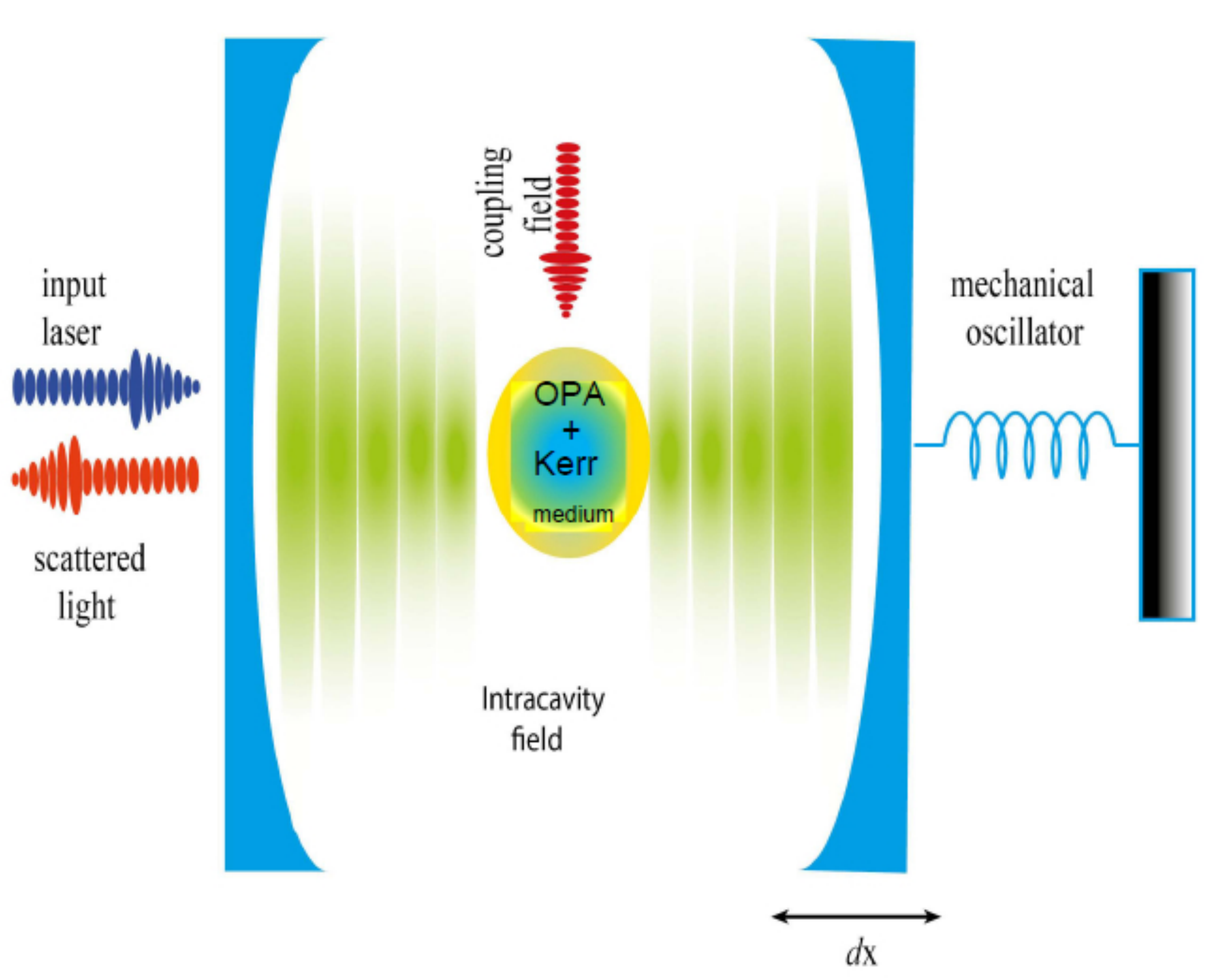}
\caption{
 (Color online) Schematic picture of the setup studied in the text. The cavity contains a Kerr-down conversion system which is pumped by a coupling field to produce parametric oscillation and induce Kerr nonlinearity in the cavity.}
\label{fig:fig1}
\end{figure}

As is shown in Fig. \ref{fig:fig1}, we consider a Kerr-down conversion optomechanical system composed of a  degenerate OPA and a nonlinear Kerr medium placed within a Fabry-Perot cavity formed by a fixed partially transmitting mirror and one movable perfectly reflecting mirror  in equilibrium with a thermal bath at  temperature  $ T_0$. The movable mirror  is free to move along the cavity axis and is treated as a quantum mechanical harmonic oscillator with effective mass $m$, frequency $\omega_{m}$ and energy decay rate $\gamma_{m}=\omega_{m}/Q$ where $Q$ is the mechanical quality factor. The cavity field is coherently driven by an input laser field with frequency $\omega_{L}$ and amplitude $\varepsilon$ through the fixed mirror. Furthermore, the system is pumped by a coupling field to produce parametric oscillation and induce the Kerr nonlinearity in the cavity. In our investigation, we restrict the model to the case of single-cavity and mechanical modes\cite{law, mancini-singlemode, bose}. The single cavity-mode assumption is justified in the adiabatic limit, i.e., $\omega_{m}\ll\pi c/L$ in which $c$ is the speed of light in vacuum and $L$ is the cavity length in the absence of the cavity field. We also assume that the induced resonance frequency shift of the cavity and  the Kerr medium are much smaller than the longitudinal-mode spacing in the cavity. Furthermore, one can restrict to a single mechanical mode when the detection bandwidth is chosen such that it includes only a single, isolated, mechanical resonance and mode-mode coupling is negligible. 

Under these conditions, the total  Hamiltonian of the system  in a frame  rotating at the laser frequency $\omega_{L}$ can be written as
\begin{equation} \label{H}
H=H_{0}+H_{1}, 
 \end{equation}
where
\begin{subequations}\label{hamil}
\begin{eqnarray}
H_{0}&=&\hbar(\omega_{0} -\omega_{L}  ) a^{\dagger} a+ \frac{ \hbar \omega_m} {2} (  q^2+p^2) -\hbar g_{0} a^{\dagger} a q\nonumber\\
&&+
i\hbar\varepsilon(a^{\dagger}-a),\\  
H_{1}&=&i \hbar G( e^ { i\theta} a^{\dagger 2}  -e^ {-i\theta} a^2 )+\hbar\chi a^{\dagger 2} a^{2} .
\end{eqnarray}
\end{subequations} 
The first two terms in $H_{0} $ are, respectively, the free Hamiltonians of the cavity field with annihilation(creation) operator   $a(a^{\dagger})$  and the movable mirror with resonance frequency $\omega_{m}$ and dimensionless position and momentum operators $q$ and $p$. The third term describes the  optomechanical coupling between the cavity field and the mechanical oscillator due to the  radiation pressure force with coupling constant $g_0=\dfrac{\omega_0}{L}\sqrt{\dfrac{h}{m \omega_m}}$,  and the last term in $H_{0} $  describes the driving of the intracavity mode with the input laser.   Also, the two terms in  $H_{1} $ describe, respectively, the coupling of the intracavity field with  the  OPA and the Kerr medium; $G$ is the nonlinear gain of the OPA which is proportional to the pump power driving amplitude, $\theta$ is the phase of the field driving the OPA, and $\chi$ is the anharmonicity parameter proportional to the  third order nonlinear susceptibility $\chi^{(3)}$  of the Kerr medium.
The input laser field populates the intracavity mode through the partially transmitting mirror, then the photons in the cavity will exert a radiation pressure force on the movable mirror.  In a realistic treatment of the dynamics of the system , the cavity-field damping due to the  photon-leakage through the incomplete mirror and the Brownian noise associated with the coupling of the oscillating mirror to its thermal bath should be considered. Using the input-output formalism of quantum optics\cite{gardiner}, we can consider the effects of these sources of noise and dissipation in the quantum Langevin equations of motion. For the given Hamiltonian (\ref{H}), we obtain the following nonlinear equations of motion
\begin{subequations}\label{langevin}
\begin{eqnarray}
\dot{q}&=& \omega_m p,\\
\dot {p}&=&-\omega_m q + g_{0}a^{\dagger}a-\gamma_{m} p+\xi ,\\
\dot{a}&=&-i (\omega_{0}-\omega_{L})a+i g_{0} q a +\varepsilon -2 i \chi a^{\dagger}a^{2}   \nonumber\\ 
 &&+
2G a^{\dagger}e^{i\theta} -\kappa a +\sqrt{2\kappa}a_{in},
\end{eqnarray}
\end{subequations}
where $\kappa$ is the cavity  decay rate through the input mirror and  $a_{in}$ is the input vacuum noise operator  characterized by the following correlation functions  \cite{gardiner}
 \begin{subequations}\label{noise1}
\begin{eqnarray}
< a_{in}(t) a_{in}^{\dagger}(t')>&=& \delta(t-t'),\\
< a_{in}(t)a_{in} (t')>&=&<a_{in}^{\dagger} (t) a_{in}^{\dagger}(t')>=0 .
\end{eqnarray}
\end{subequations}
The Brownian noise operator $\xi$ describes the heating of the mirror by the thermal bath at temperature $T_0$ and is characterized by the following correlation function  \cite{gardiner, giovannetti, landau}
\begin{equation}\label{noise2}
<\xi(t)\xi(t')>=\frac{\hbar \gamma_{m}m}{2\pi} \int \omega e^{-i\omega(t-t')}[\coth(\frac{\hbar\omega}{2k_{B} T_0} )+1]d\omega,
\end{equation}
where $k_{B} $ is the Boltzmann constant. 

We are interested in the  steady-state regime and  small fluctuations with respect to the steady state. Thus we obtain the steady-state mean values of $p$, $q$ and $a$ as
  \begin{equation}\label{qs}
 p_{s}=0, \quad q_{s}=\frac{ g_{0}}{ \omega_{m}}a_{s}^{2},
 \end{equation}
 \begin{equation}\label{as}
a_{s}=\frac{\varepsilon}{\sqrt{(\Delta-2 G \sin \theta)^2+\kappa_{-}^2}},
\end{equation}
where  $q_{s}$ denotes  the new equilibrium position of the movable mirror, $\kappa_{-}=\kappa-2G \cos\theta$ , and $\Delta=\omega_{0}-\omega_{L}-g_{0} q_{s} +2\chi a_{s}^{2}= \Delta_0+(2\chi-g_0^2/\omega_m)a_{s}^{2}$ is the effective detuning of the cavity which includes both  the radiation pressure  and the  Kerr medium effects. It is obvious that the optical path and hence the cavity detuning are modified in an intensity-dependent way. The first modification which is a mechanical nonlinearity, arises from the  radiation pressure-induced coupling between the movable mirror and the cavity field and the second modification  comes from the presence of the nonlinear Kerr medium in the optomechanical system. Since the mean intracavity photon number in the steady state $I_a(=a_s^2 )$ satisfies a third-order equation, it  can have three real solutions and hence the system may exhibit multistability for a certain range of parameters.  The multisolution region exists for intracavity intensity values between $I_{-}$ and $I_{+}$ with
\begin{equation}
I_{\pm}=\dfrac{(4G \sin\theta-2\Delta_0)\pm \sqrt{(2G \sin\theta-\Delta_0)^2-3\kappa_{-}^2}}{3(2\chi-g_0^2/\omega_m)}.
\end{equation}
The multistability of the solution fails if   $\chi= g_0^2/2\omega_m$ and   requires  $\vert2G \sin\theta-\Delta_0\vert>\sqrt{3}\kappa_{-}$.

\begin{figure}[ht]
\centering
\includegraphics[width=2.8 in]{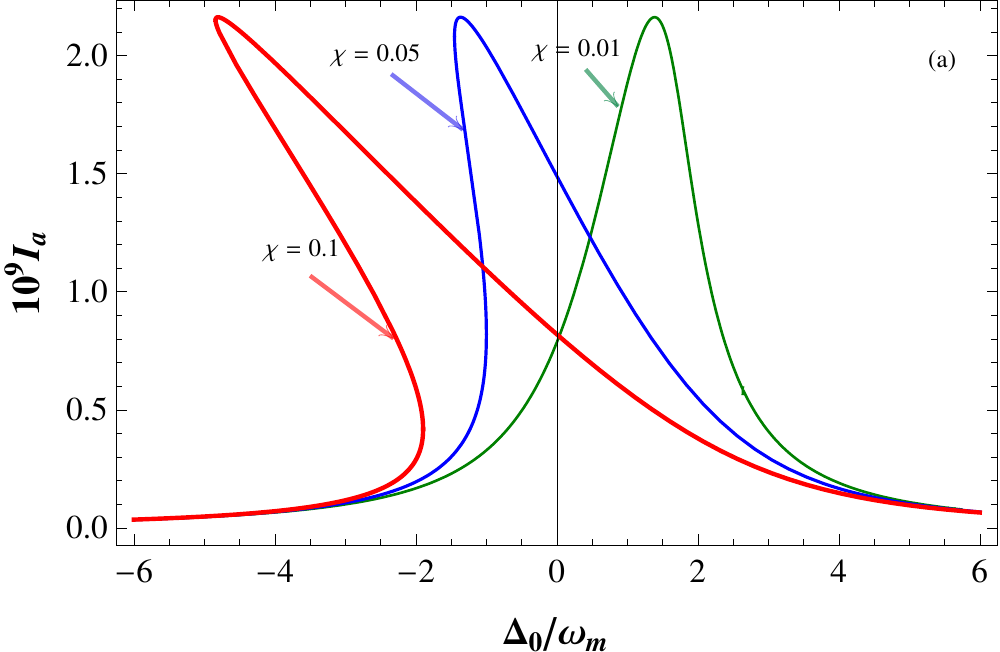}
\includegraphics[width=2.8 in]{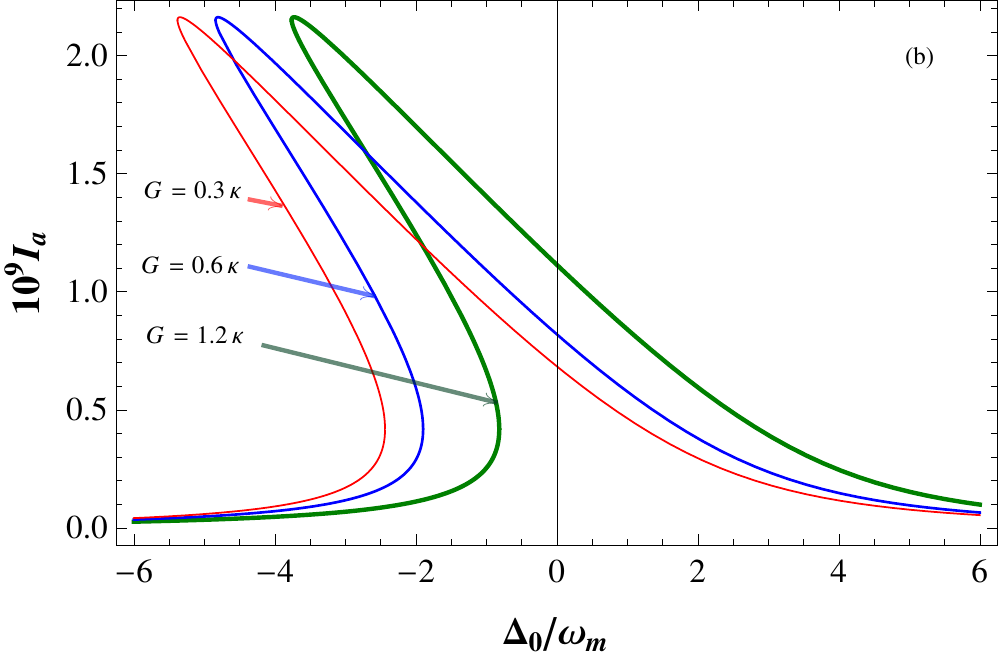}
\includegraphics[width=2.8 in]{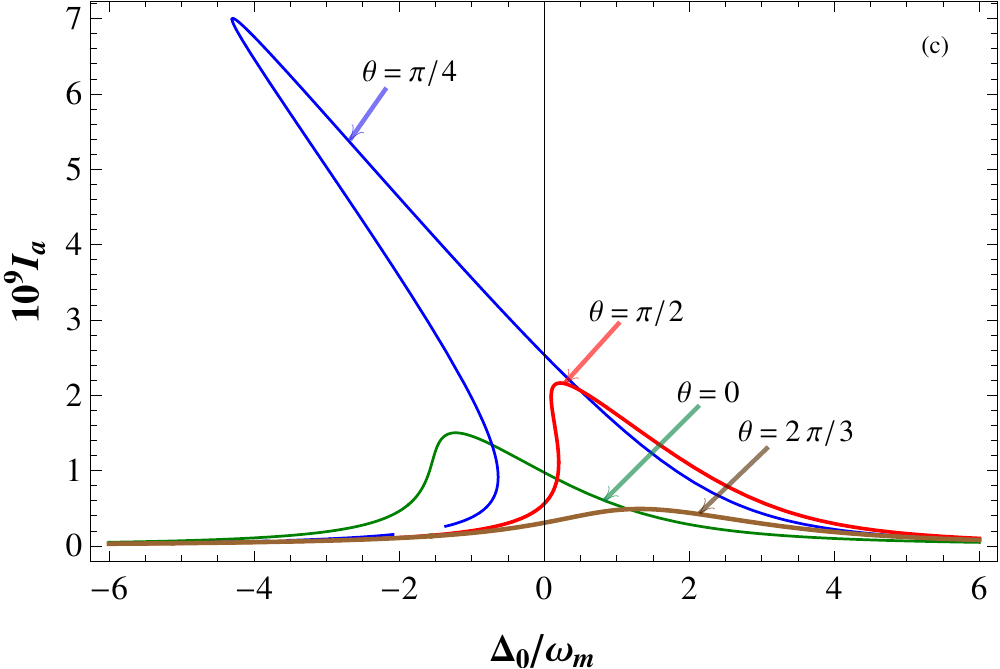}
\caption{ (Color online)The mean intracavity photon number as a function of normalized bare detuning $\Delta_0/\omega_m$ : (a) for different values of the anharmonicity parameter$\chi$ with $G=0.6\kappa$ and $\theta=\pi/2$, (b) for different values of the parametric nonlinearity $G$ with $\chi=0.1$ Hz and $\theta=\pi/2$, and  (c) for different values of $\theta$ with $G=1.1\kappa$ and $\chi=0.04$ Hz.
}
\label{fig:as}
\end{figure}

To study the effect of the presence of the Kerr-down conversion nonlinearity   on the  steady-state response of the optomechanical system  we  consider a  cavity with length $L=1$mm and decay rate $\kappa= 0.9\omega_m$ which is driven by  a laser with $\lambda=810$nm. The      mass,  the mechanical  frequency  and  the damping rate  of the  oscillating  end mirror are $m=5$ ng, $\omega_m/2\pi=10$MHz,  $\gamma_m=100$ Hz  and the environment temperature  is $T_0 = 400$mK. This  set of parameters  is close to several optomechanical experiments\cite{gigan,klechner1,klechner2,carmon}.    In Fig. \ref{fig:as} we plot the mean intracavity photon number as a function of the bare detuning $\Delta_0$ for input power $P=15$mW for various values of $\chi$  (Fig. \ref{fig:as}a) , $G$ (Fig. \ref{fig:as}b), and $\theta$(Fig. \ref{fig:as}c). Figure \ref {fig:as}a shows that for $\chi<g_0^2/2\omega_m$ ($\chi=0.01$) the presence of both intrinsically optomechanical  and Kerr nonlinearities shift the center of the resonance while the curve is nearly Lorentzian. However, for $\chi>g_0^2/2\omega_m$ the resonance frequency of the cavity shifts to the  lower values and    the  third order polynomial equation for $I_a$ has   three real roots for  $\Delta_0<0$. The frequency shift of the cavity mode  and multi solutions due to the  Kerr nonlinearity can be reduced or compensated by the term  $2G \sin\theta$  which  acts to shift the cavity resonant frequency to the right for  $\theta\geq\pi/2$(Figs. \ref{fig:as}b,c).

 \section{Dynamics Of Small Fluctuations}\label{sec2}
In order to investigate the dynamics of the system, we need to study the dynamics of small fluctuations near the steady state. We assume that the nonlinearity in the system is weak and decompose  each operator in Eq.(\ref{langevin}) as the sum of its steady-state value and a small fluctuation with zero mean value,
\begin{equation}\label{linear}
a=a_{s}+\delta a,\quad q=q_{s}+\delta q,\quad p=p_{s}+\delta p.
\end{equation}
Inserting the above linearized forms of the system operators into Eq.(\ref{langevin}), and defining the cavity field quadratures $\delta x=(\delta a+\delta a^{\dagger})/\sqrt{2} $ and $\delta y=i(\delta a^{\dagger}-\delta a)/\sqrt{2} $ and the input noise quadratures $\delta x_{in}=(\delta a_{in} +\delta a^{\dagger}_{in} )/\sqrt{2} $ and $\delta y_{in} =i(\delta a^{\dagger}_{in} -\delta a_{in}) /\sqrt{2}$   the linearized quantum Langevin equations  for the  fluctuation operators   can be written in the compact matrix form
 \begin{equation}\label{matrixform}
 \dot{u}=M u(t)+n(t),
 \end{equation}
where $u(t) =(\delta q,\delta p,\delta x,\delta y)^{T}$ is the vector of fluctuations, $n(t) =(0,\xi ,\sqrt{2\kappa}\delta x_{in},\sqrt{2\kappa} \delta y_{in})^{T}$ is the vector of noise sources and the  matrix $M$ is given by
\begin{equation}\label{M}
M=\left(
\begin{array}{cccc}
0 & \omega_m & 0 & 0\\ 
-\omega_m & -\gamma_{m} & g_1& 0 \\ 
0 & 0 &  -\kappa+\Gamma_r & \Delta_{1}+\Gamma_i  \\ 
g_1 & 0 & -\Delta_{1}+\Gamma_i   &  -\kappa-\Gamma_r
\end{array} \right)
\end{equation}
where  $g_1=\sqrt{2}g_0 a_s$ is the enhanced optomechanical coupling rate,    $\Gamma_r$ ($\Gamma_i$ ) is  the real (imaginary) part of $\Gamma=2G e^{i\theta}-2 i \chi a_s^2$,  and   $\Delta_{1}=\Delta+2\chi a_s^2$ .
\subsection{Stability analysis of the  solutions}
Here, we concentrate on the stationary properties of the system. For this purpose, we should consider the steady-state condition governed by Eq.(\ref{matrixform}). The system is stable only if the real part of all eigenvalues of the matrix $M$ are negative, which is also the requirement of the validity of the linearized method. 
The parameter region in which the system is stable can be obtained from the Routh-Hurwitz criterion\cite{routh}, which gives the following three independent conditions:
\begin{subequations}\label{routh-1-3}
\begin{eqnarray}
s_{1}&=&\kappa^2+\Delta_1^2-\vert\Gamma\vert^2> 0,\\
s_{2}&=&\omega_m(\kappa^2+\Delta_1^2-\vert\Gamma\vert^2)-g_0^2(\Delta_1+\Gamma_i)>0,\\
s_{3}&=&2\kappa\gamma_m\lbrace(s_1-\omega_m^2)^2+ (\gamma_m+2\kappa)(\gamma_m s_1+2\kappa\omega_m^2)\rbrace\nonumber\\
&&+g_0^2(\Delta_1+\Gamma_i)\omega_m(2\kappa+\gamma_m)^2  >  0.
\end{eqnarray}
\end{subequations}
The violation of the third  condition, $s_3 < 0$, indicates instability in the region $\Delta_1+\Gamma_i<0$. For the bare cavity ($G=\chi=0$) this condition reduces to the instability in the domain of  blue-detuned laser. Within this frequency range, the effective mechanical damping rate becomes negative  and self-sustained  oscillations set in\cite{marquardt, ludwig-ss}.
The violation of the second  condition, $s_2 < 0$, indicates instability in the region $\Delta_1+\Gamma_i>0$. For a bare cavity   this condition cause a bistability of the system\cite{dorsel}. There is an additional stability condition given by $s_1$, which is always satisfied for the bare cavity, and gives the condition for the threshold for parametric oscillation. Accordingly, for $\Delta_1+\Gamma_i>0$ we can define the following stability parameters 
\begin{eqnarray}\label{eta1,2}
\eta_1&=&1-\dfrac{g_1^2(\Delta_1+\Gamma_i)}{\kappa^2+\Delta_1^2-\vert\Gamma\vert^2},\\
\eta_2&=&=1-\dfrac{\vert\Gamma\vert^2}{\kappa^2+\Delta_1^2}.
\end{eqnarray}
 For $G=\chi=0$, $\eta_1$  reduces to the well known "bistability parameter" \cite{genes-bistability}.

\begin{figure}[ht]
\centering
\includegraphics[width=2.8 in]{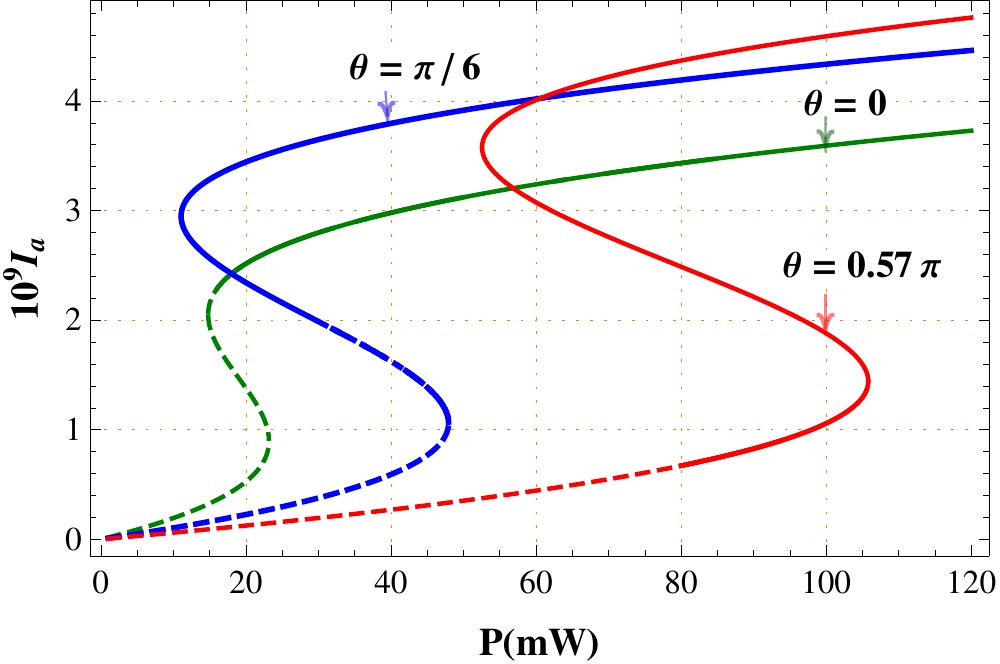}
\caption{Mean photon number $I_a$ versus the input power $P$ at $\Delta_0=-2.5\omega_m$.The solid and dashed lines correspond to the stable and unstable branches, respectively. The parameters are: $G=\kappa$, $\chi=0.05$ Hz, $\kappa=0.9\omega_m$. Other parameters are
the same as those in Fig.  \ref{fig:as} 
}
\label{fig:tristability}
\end{figure}

One of the main features arising from the  Kerr-down conversion nonlinearity is the appearance of  three stable states for the mirror.  Figure  \ref{fig:tristability} shows the hystersis loop for the intracavity mean photon number when  $\Delta_0<0$ (blue-detuned laser).  In this figure  the  unstable solutions  are represented by  dashed lines. It shows that depending on the value of $\theta$ the steady-state response of the mirror can be monostable, bistable and tristable. 
From an experimental point of view,  controllable triple-state switching is possible practically   by adding a pulse sequence onto the input field\cite{min-xiao}. Such tristability   can be used for all-optical switching purposes function as  memory devices for optical computing and quantum information processing.
\subsection{Correlation matrix of the quantum fluctuations of the system}
Due to the linearization method and the Gaussian nature of the noise operators the asymptotic steady state of the quantum fluctuations  is a zero mean Gaussian state. 
 As a consequence,   the steady state can be fully  characterized by the  covariance matrix (CM)$V$. This formalism provides a unified framework for exploring  both cooling of the mechanical oscillator  and phonon-photon entanglement. 

When the stability conditions of Eq.(\ref{routh-1-3}) are fulfilled, we can solve   Eq.(\ref{matrixform}) for the  $4\times4$ stationary correlation  matrix  $V$\cite{vitali-CM, genes-CM, mari}
\begin{equation}\label{Lyapunov}
M V+V M^{T}=-D,
\end{equation}
where the elements of the correlation matrix $V$ are defined as $  V_{i j} = \left\langle  u_{i}u_{j}+u_{j}u_{i}\right\rangle/2$ and $D=$Diag$[0,\gamma_{m}(2\bar{n} +1),\kappa,\kappa]$   is the diagonal diffusion matrix in which we used the  following approximation 
 \begin{equation}
 \omega \coth(\frac{\hbar\omega}{2 k_{B}T_0})\simeq \omega_{m}\frac{2 k_{B}T_0}{\hbar\omega_{m}}\simeq \omega_{m}(2\bar{n}+1),
 \end{equation}
 where $\bar{n}=[e^{\hbar\omega_{m}/k_{B}T_0}-1]^{-1}$. The Lyapunov Eq. (\ref{Lyapunov}) for the steady state CM  can be straightforwardly solved. However, the general exact expression is too cumbersome and will not be reported here.
\section{Effects of the Kerr-down conversion nonlinearity on ground state cooling and stationary entanglement}
It has been shown \cite{genes-CM, marq,wilson, Paternostro, dob, gen} that within  each optomechanical cavity the radiation pressure coupling can lead the  system into a stationary state  with genuine quantum features, including  cooling of the resonator towards the ground state  and     photon-phonon entanglement. It is therefore interesting to investigate these quantum features in the presence of Kerr-down conversion nonlinearity. 

The CM , $V$, contains all the information about the steady state.  In particular, the mean energy of the mechanical oscillator is given by
\begin{equation}
U=\frac{\hbar\omega_m}{2} (<\delta q>^2+<\delta p>^2)=\hbar\omega_m(n_{eff}+\dfrac{1}{2}),
\end{equation}
where $n_{eff}$ is the mean effective excitation number of the mechanical mode corresponding to an effective mode temperature $T=\hbar\omega_m/(k_B ln(1+1/n_{eff}))$. The ground-state cooling is approached if $n_{eff} \simeq0$,  or $U\simeq \hbar\omega_m/2$.

Moreover, the photon-phonon entanglement   can be quantified by the logarithmic negativity $E_{N}$ \cite{adesso} 
 \begin{equation}
E_N=\mathrm{max}[0,-\mathrm{ln} 2 \eta^-],
 \end{equation}
where  $\eta^{-} \equiv2^{-1/2}\left[\Sigma(V)-\sqrt{\Sigma(V)^2-4 \mathrm{det} V}\right]^{1/2}$ is the lowest symplectic eigenvalue of the partial transpose of the  CM  with $\Sigma(V)=\mathrm{det} V_{A} +\mathrm{det}V_{B} -2\mathrm{det} V_{C}$, and we have used the $2\times 2$ block form of the CM 
   \begin{equation}
 \left( \begin{matrix}
 V_{A} &V_{C}  \\ 
V_{C}^{T} & V_{B} 
\end{matrix} \right),
  \end{equation}   
 with $V_{A}$ associated to the oscillating mirror, $V_{B}$  to the cavity mode, and $V_{C}$ describing the optomechanical correlations.
 
\begin{figure}[ht]
\centering
\includegraphics[width=2.8 in]{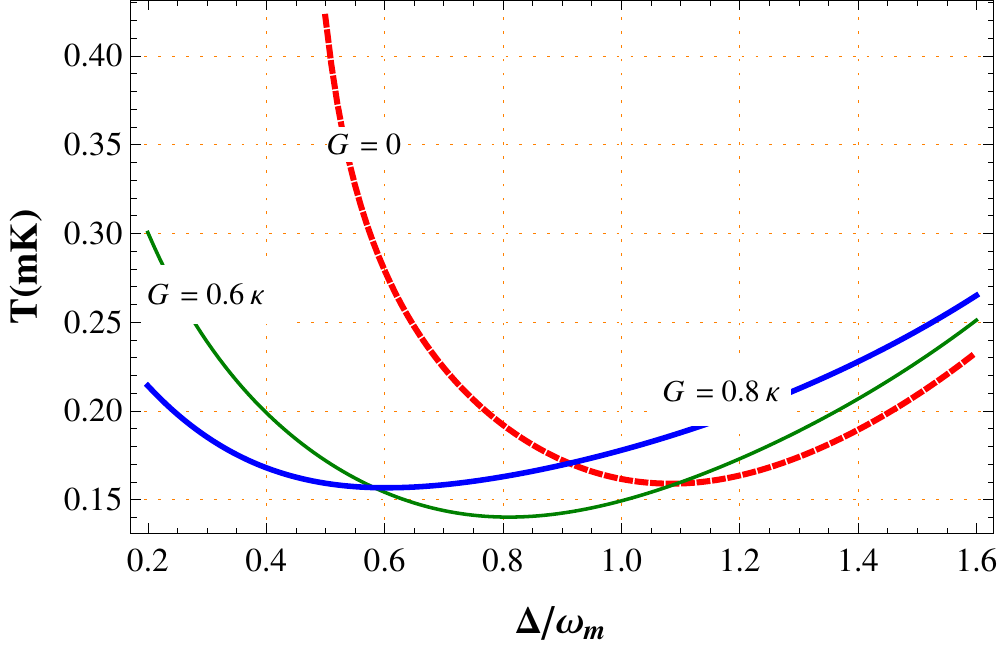}
\caption{(Color online) Plot of  the effective temperature $T$  versus $\Delta/\omega_m$  for different values of $G$. The red dashed curve corresponds to the bare cavity. The parameters are :$\chi=0.05$Hz, $\theta=0.81\pi$ for $G=0.6\kappa$ and $\theta=0.71\pi$ for $G=0.8\kappa$, $T_0=400$mK,  $P=5$ mW, and  $\kappa=0.3\omega_m$.  Other parameters are the same as those in Fig.  \ref{fig:tristability}. 
}
\label{fig:t1}
\end{figure}

We now focus on  numerical examples  to see  how the Kerr-down conversion nonlinearity affects the ground-state cooling of the movable mirror. Figure \ref{fig:t1} shows the variation of the effective  temperature $T$  with $\Delta/\omega_{m}$  and   $G$  for  the red-detuned laser($\Delta_0>0$ ) and in the monostable regime. In  this figure we have assumed  the good-cavity limit, $\kappa=0.3\omega_m$, and the reservoir temperature to be  $T_0 = 0.4 K$.  The effective mode temperature of  the bare cavity (red dashed line) is plotted for comparison.  It shows that the parametric process inside the cavity leads to lower temperature. To illustrate  this explicitly, we can  solve Eq. (\ref{Lyapunov}) in the limit of large mechanical quality factor and low-temperature environment (i.e., $\omega_m\gg\gamma_m$ and $\kappa\gg \bar{n}\gamma_m$) and obtain the position ($V_{11}$)  and momentum ($V_{22}$) variances and $n_{eff}$ by  using $n_{eff}=\frac{1}{2}(V_{11}+V_{22}-1)$. The results are as follows
\begin{eqnarray}
V_{11}&=&\frac{\eta_1\omega_m^2+(\Delta_1+\Gamma_i)^2+(\kappa+\Gamma_r)^2}{4 \eta_1\omega_m(\Delta_1+\Gamma_i)},\\
V_{22}&=&\frac{\omega_m^2+(\Delta_1+\Gamma_i)^2+(\kappa+\Gamma_r)^2}{4 \omega_m(\Delta_1+\Gamma_i)},
\end{eqnarray}
and for $\eta_1\simeq1$,
\begin{equation}
n_{eff}=\dfrac{(\Delta_1+\Gamma_i-\omega_m)^2+\kappa_+^2}{4\omega_m(\Delta_1+\Gamma_i)},
\end{equation}
where $\kappa_+=\kappa+\Gamma_r$.  For a bare cavity $n_{eff}$ reduces  to  Eq. (6) in \cite{wilson}. The minimum attainable phonon number can be achieved for   $\Delta=-2G\sin\theta+\sqrt{\omega_m^2+\kappa_+^2}$,
\begin{equation}\label{nmin}
n_{min}=\dfrac{1}{2}(\dfrac{\sqrt{\omega_m^2+\kappa_+^2}}{\omega_m}-1),
\end{equation}
in agreement with  Eq. (7) in \cite{wilson} for $G=0$. As can be seen,  when  $\pi/2<\theta<3\pi/2$ the lower bound of the resolved sideband regime($\omega_m\gg\kappa_{+}$) and  $n_{min}(\simeq\kappa_+^2/4\omega_m)$ can be less than that of  a bare cavity.  For the data used in Fig. \ref {fig:t1}, and for   $G=0.6\kappa$,  $\theta=0.81\pi$   we have $\kappa_+\simeq10^{-6}\kappa$  and the minimum effective temperature is  about   $0.14$mK at $\Delta\simeq0.8\omega_m$. Also, for  $G=0.8\kappa$ and $\theta=0.71\pi$ we have  $\kappa_+\simeq 8\times10^{-6}\kappa$ and the minimum effective temperature is about   $0.16$mK at $\Delta\simeq0.6\omega_m$.  

\begin{figure}[ht]
\centering
\includegraphics[width=2.8 in]{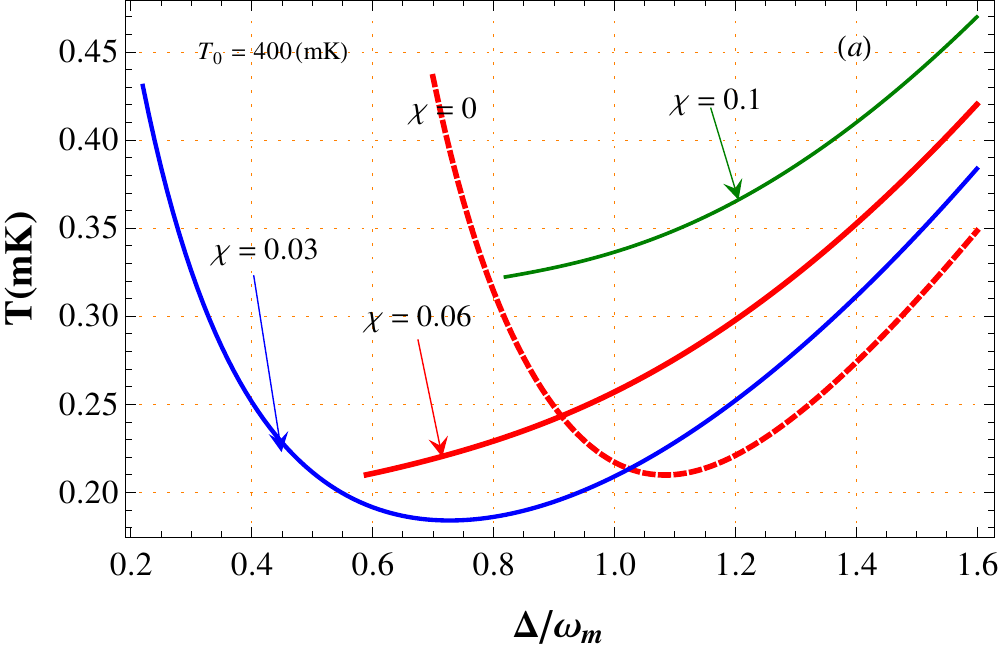}
\includegraphics[width=2.8 in]{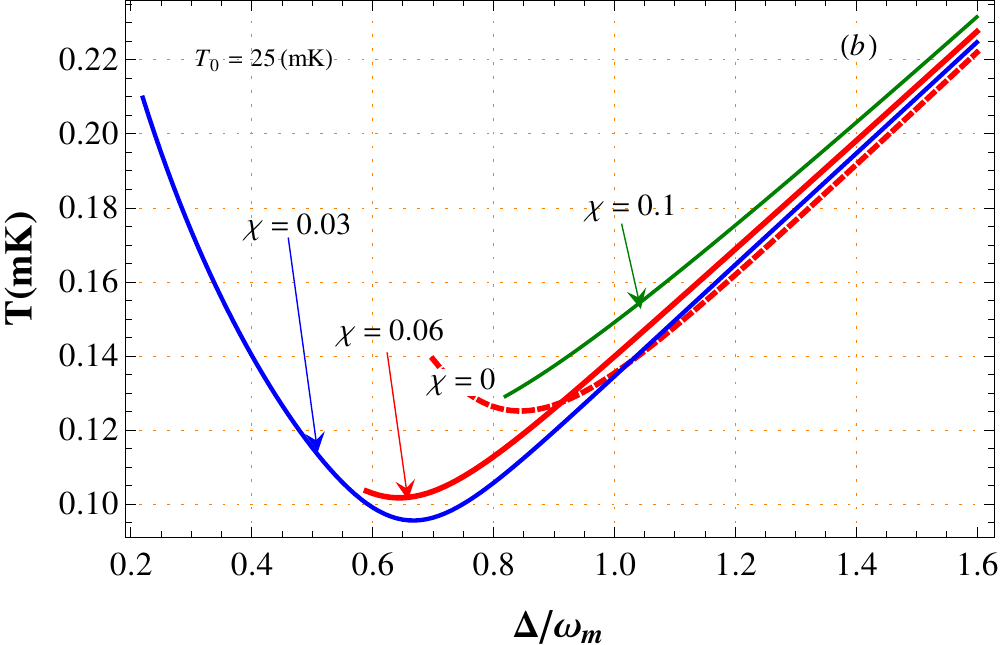}
\caption{(Color online) Plot of  the effective temperature $T$  versus $\Delta$  for environment temperature (a) $T_0=400$(mK) and (b) $T_0=25$(mK) for different values of $\chi$.The red dashed curve corresponds to the  cavity containing only the  gain nonlinearity. The parameters are:$G=0.8\kappa$,  $\theta=3\pi/4$, and  $P=5$ mW.  Other parameters are
the same as those in Fig.  \ref{fig:tristability} and the stability parameter is fixed to be $\eta_1=0.99$.
}
\label{fig:t-kerr}
\end{figure}

According to Eq. (\ref{nmin}) it seems that $n_{min}$ does not depend on the anharmonicity parameter $\chi$,  but  it should be noted  that  even if  the effect of Kerr nonlinearity on  $\eta_1$ can be  eliminated, the above analysis is carried out for the limit of low-temperature environment ($\kappa\gg \bar{n}\gamma_m$). In a realistic case the thermal noise  will also be present, which in turn modifies the minimum attainable temperature. Figures \ref{fig:t-kerr}(a ) and (b)  show the variation of effective temperature with $\Delta$ for different values of $\chi$  and for two different environment temperatures $T_0=400$mK and  $25$mK. For  $G=0.8\kappa$ and $\theta=3\pi/4$    the minimum temperature is achieved for the optimal detuning  $\Delta\simeq0.66\omega_m$.  As can be seen, the system is unstable in this range   for  $\chi=0$ and $\chi=0.1$.  For $\chi=0.03$ and $\chi=0.06$ the stability domain is extended to the desired  range and the effective temperature is increased with increasing $\chi$. As shown in Fig. \ref{fig:t-kerr}(b) this heating effect is smaller for lower environment temperature  $T_0=25$mK.  
\begin{figure}[ht]
\centering
\includegraphics[width=2.8 in]{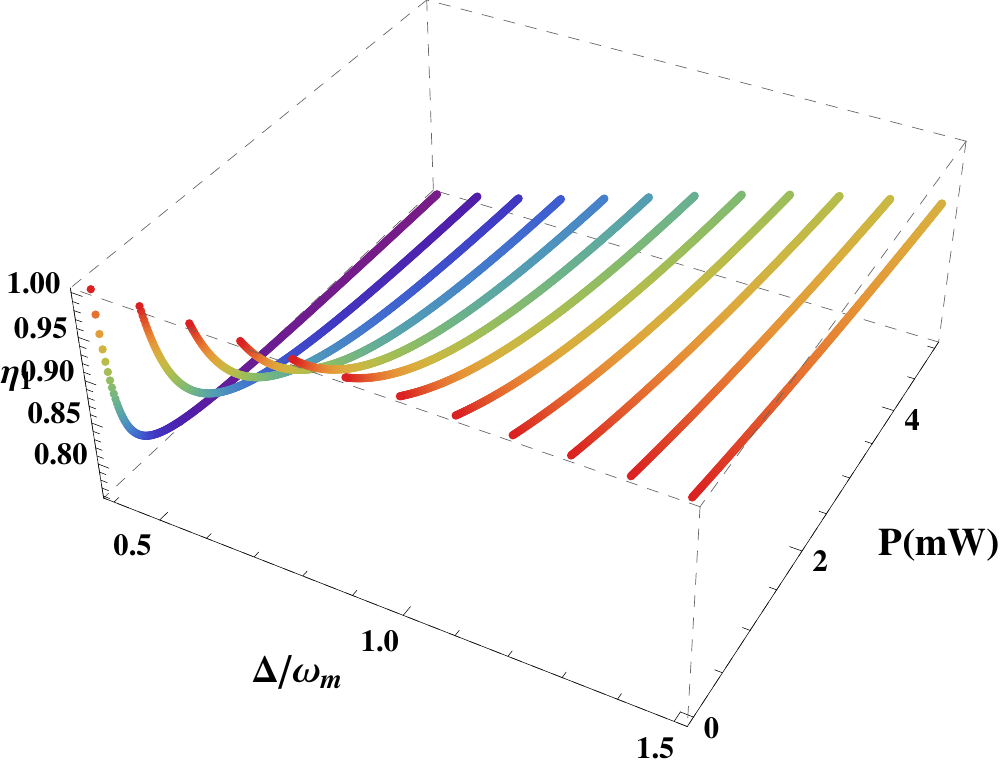}
\caption{(Color online) Plot of  the stability parameter $\eta_1$  versus $\Delta/\omega_m$  and input power $P$  for $\chi=0.03$.  Other parameters are the same as those in Fig.  \ref{fig:t-kerr}(a).
}
\label{fig:eta1}
\end{figure}
\begin{figure}
\centering
\includegraphics[width=2.5 in]{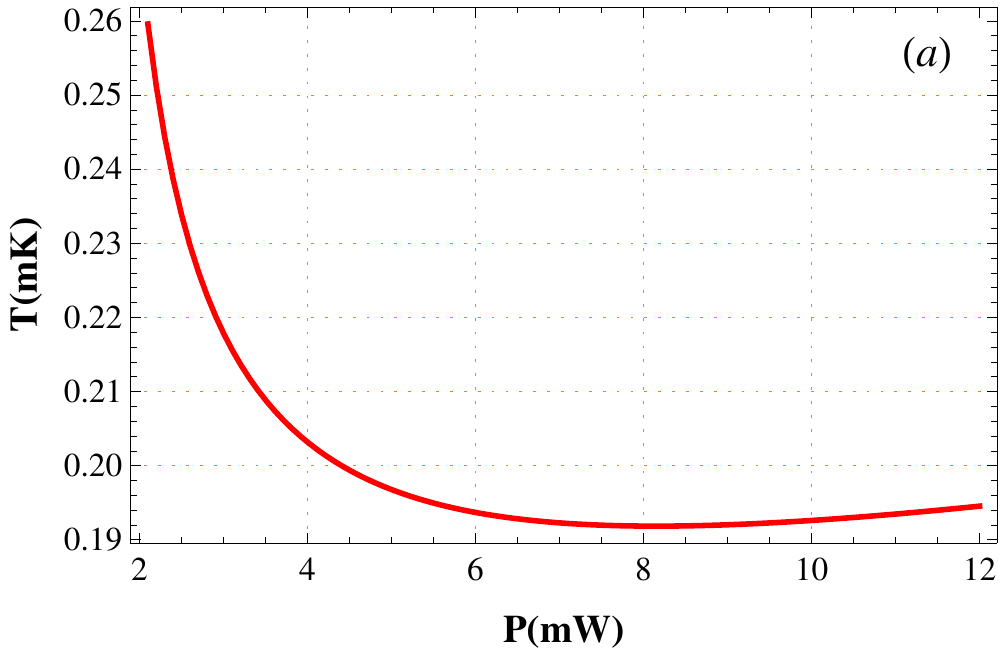}
\includegraphics[width=2.5 in]{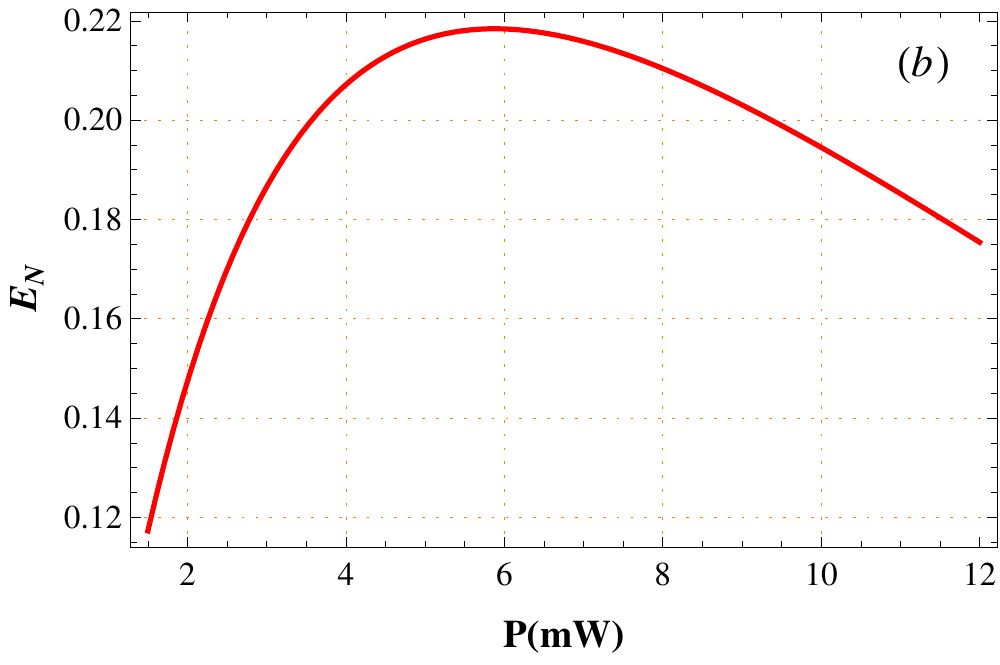}
\caption{(Color online) Plot of  (a) the effective temperature $T$  and (b) the logarithmic negativity versus the input power $P$  for $\Delta=0.5\omega_m$, $G=1.3\kappa$, $\theta=0.67\pi$ and  $\chi=0.05$.  Other parameters are
the same as those in Fig.  \ref{fig:t1}.
}
\label{fig:ent1}
\end{figure}

Also, in   Fig. \ref{fig:t-kerr}  the dependence  of effective temperature on the Kerr nonlinearity has been illustrated   for fixed values of  stability parameter $\eta_1$. It should be emphasized, however, that  the controllable ground state cooling of the vibrational mode  is possible only  if the limit  $\eta_1\simeq1$ is reachable in the presence of the nonlinear medium.  Figure  \ref{fig:eta1} shows the variation of the parameter $\eta_1$ versus the input power $P$ and the  normalized effective detuning $\Delta/\omega_m$ for the data of Fig.  \ref{fig:t-kerr}(a ) and for $\chi=0.03$. As can be seen, $0.8<\eta_1<1$ and the  required condition   holds actually very well. Also, Fig. \ref{fig:eta1} shows that  for a fixed effective detuning, in contrast to   the  case of bare cavity  \cite{ghobadi},   $\eta_1$ is not a linear function of the input power $P$. This feature arises from the Kerr nonlinearity (the term $2\chi a_s^2$ in $\Delta_1$ and $\Gamma$ in  Eq.(\ref {eta1,2})) and   allows to approach  significant entanglement  simultaneously with the ground state cooling  of the mirror( since one can enhance optomechanical coupling by increasing the input power while $\eta_1$ remains approximately unaffected).  As an example, Fig. \ref{fig:ent1} represents the optomechanical entanglement and the effective temperature for $G=1.3\kappa$, $\chi=0.05$ , and  $\Delta=0.5\omega_m$ as  functions of the input power.  It shows that the minimum value of the effective temperature $T$ and  the maximum  value of entanglement  is achieved in   the same range  of the input power.  

Nonetheless, we will show that as in the case of bare cavity,  entanglement and cooling are different phenomena and generally   are optimized  in  different regimes. We study the effects of $\theta$, $G$ and $\chi$ on $E_N$ separately to find the regime of maximal phonon-photon entanglement.

\begin{figure}[ht]
\centering
\includegraphics[width=2.7 in]{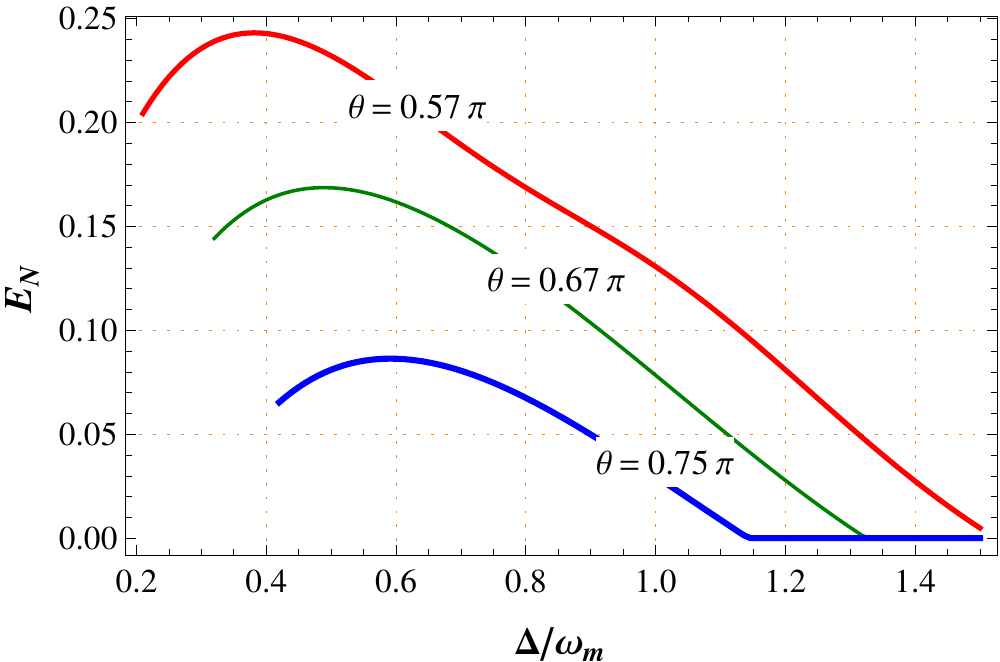}
\caption{(Color online) Plot of  the logarithmic negativity versus the normalized  effective detuning $\Delta/\omega_m$  for different values of $\theta$. The parameters are: $P=3$mW, $G=1.3\kappa$, $\chi=0.05$.  Other parameters are the same as those in Fig.  \ref{fig:t1}.
}
\label{fig:theta}
\end{figure}

We first study the behavior of $E_N$ when  the phase of the field driving the OPA, i.e., $\theta$, is varied. Figure \ref{fig:theta}  shows that entanglement increases with decreasing the phase of the field driving the OPA. This  entanglement increment is related  to the increasing of the  photon number and  the optomechanical coupling rate $g_1$ (see Fig. \ref{fig:as}(c)). 

  \begin{figure}
\centering
\includegraphics[width=2.5 in]{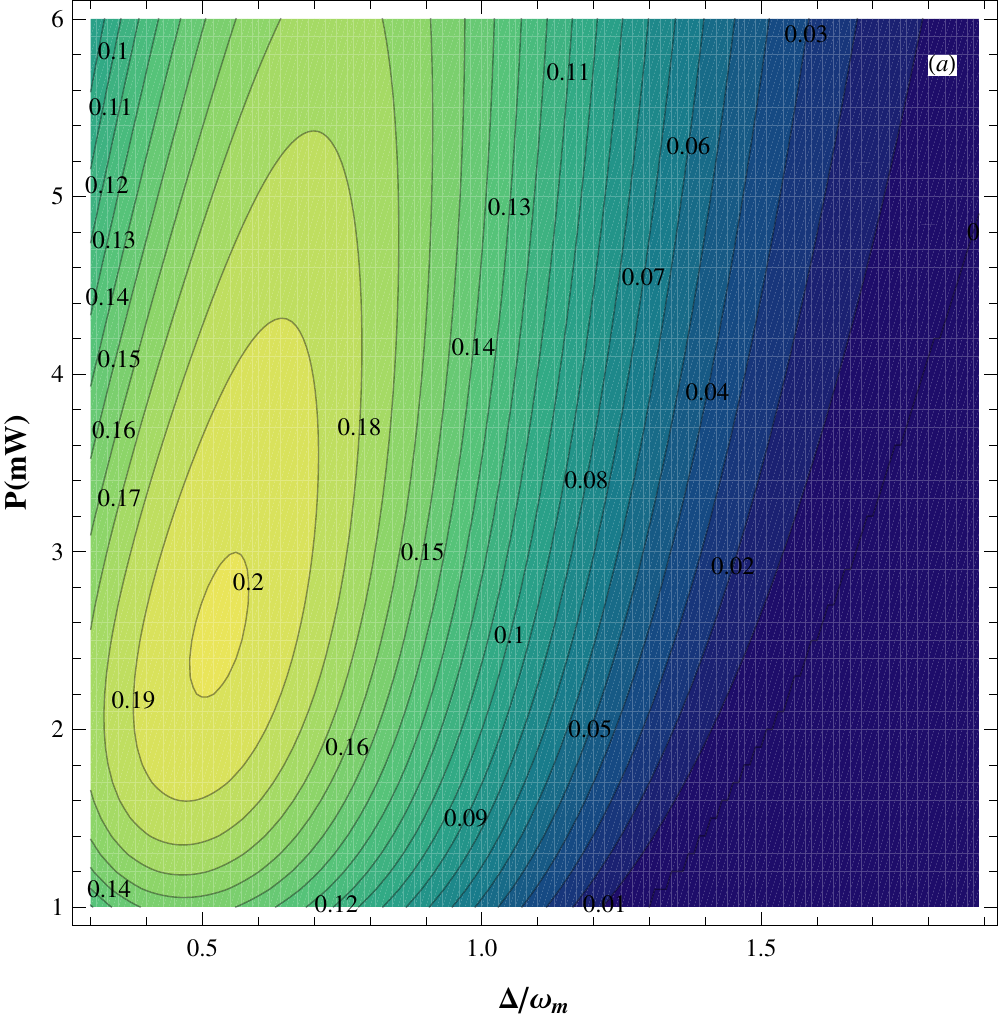}
\includegraphics[width=2.5 in]{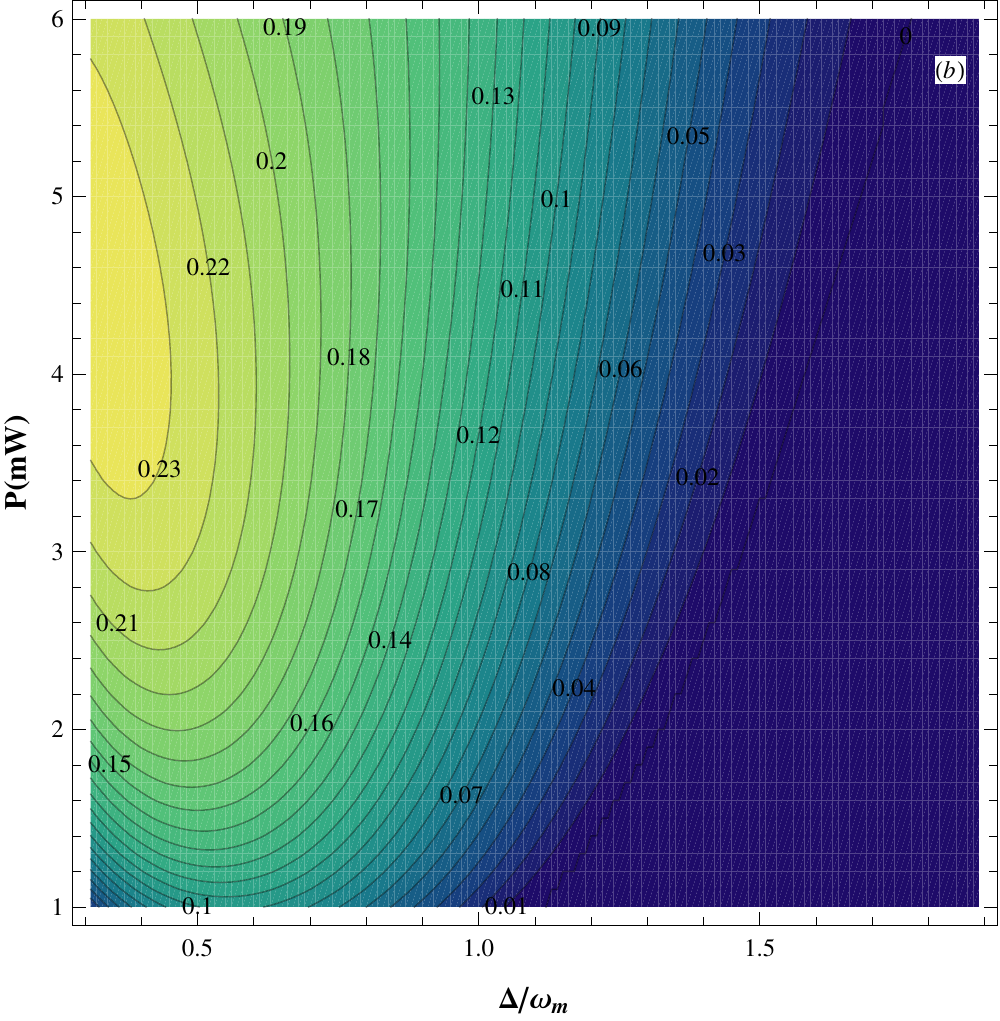}
\caption{(Color online) Contour plot of the logarithmic negativity versus the normalized  effective detuning $\Delta/\omega_m$  and input power $P$ for (a) $G=0.6\kappa$, and (b) $G=\kappa$ . The parameters are: $\theta=0.67\pi$, $\chi=0.05$.  Other parameters are the same as those in Fig.  \ref{fig:t1}.
}
\label{fig:ent-G}
\end{figure}
Figure \ref{fig:ent-G} shows the entanglement as a function  of the normalized effective  detuning $\Delta/\omega_m$ and laser power $P$ for (a)$G=0.6\kappa$ and (b) $G=\kappa$.  It shows that the maximum possible values of the logarithmic negativity occur  at low values of the input power and increase with  increasing the gain nonlinearity. It is notable that  here the entanglement increment is not  just related to the increasing  of the optomechanical coupling rate $g_1$. In Table \ref{table:1} we summarize  the maximum values of the logarithmic negativity $E_N$ and other parameters associated with it for   input powers of $P=2.5$ mW and $P=5$mW  related to  Fig. \ref{fig:ent-G}. In this table one can see that although  the optomechanical coupling strength is larger for $G=0.6\kappa$ for  both values of the input power  but $E_N$  is smaller. This feature suggests to examine   other  key parameters that determine the entanglement behavior. According to the data given in Table \ref{table:1} for both $G=0.6\kappa$ and $G=\kappa$  the  maximum values of the entanglement are obtained for $\eta_2$ very close to unity  but the effective detuning and stability parameter $\eta_1$ are smaller for $G=\kappa$. This   result is in agreement with that  obtained in Refs. \cite{genes-CM, ghobadi} for a bare cavity, showing that  the entanglement becomes maximal for $\eta_1=0$.
 It is intresting to note that in contrast to bare cavity    for $P=2.5$mW and $G=0.6\kappa$ the  maximum value of  the logarithmic negativity is obtained for the minimum value of $\eta_1$, since  at the end of  the stability region  $\Delta_1+\Gamma_i=0 $ and  $\eta_1=1$. 
 
\begin{figure}[ht]
\centering
\includegraphics[width=3 in]{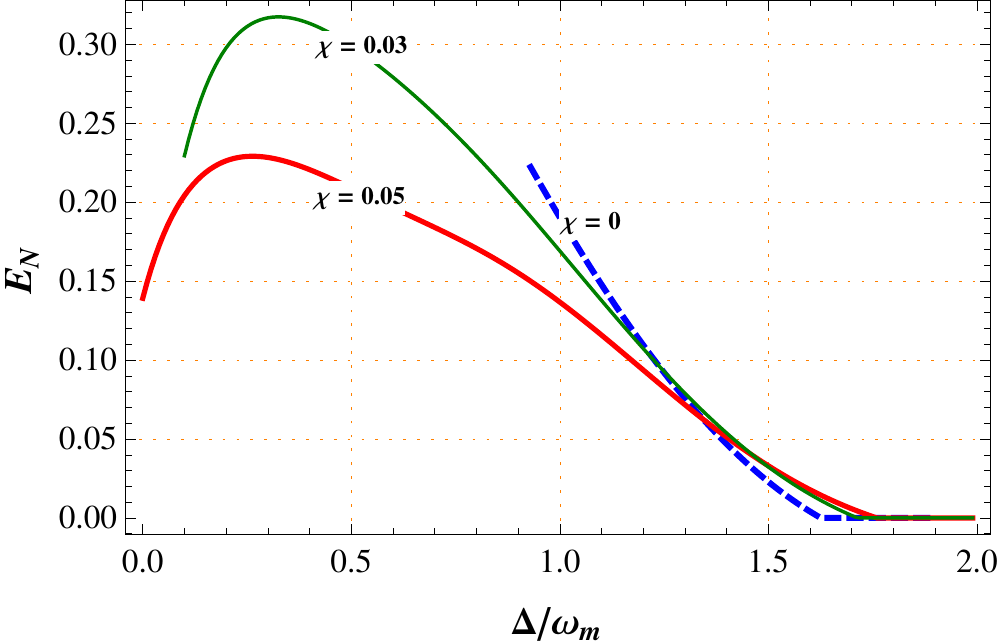}
\caption{(Color online) The logarithmic negativity versus the normalized  effective detuning $\Delta/\omega_m$  for  $\chi=0$, $\chi=0.03$, and  $\chi=0.05$. The parameters are: $\theta=0.67\pi$, $G=\kappa$, and $P=6$mW.  Other parameters are the same as those in Fig.  \ref{fig:t1}.
}
\label{fig:ent-kerr}
\end{figure}

\begin{table}
  \caption{Calculated logarithmic negativities, normalized effective detunings,  normalized  optomechanical couplings, and the two stability parameters for the  input powers $P=2.5$mW and $P=5$mW in Fig. \ref{fig:ent-G}.}
  \begin{center}
    \begin{tabular}{cccccccc}
    \hline
    $G/\kappa$ & $P(mW)$ & $E_N$ & $\Delta/\omega_m$ & $\eta_1$ &$\eta_2$ & $g_1/\omega_m$ \\
    \hline
   0.6 & 2.5 &  $0.20 $ & 0.53 & 0.77 & 0.93& 0.56  \\
    1 & 2.5 & 0.21 & 0.43 &0.67& 0.87 & 0.47  \\
     \hline
    0.6 & 5 & 0.18 & 0.68 & 0.77 & 0.87& 0.69  \\
    1 & 5 & 0.23 & 0.30 &0.66& 0.88 & 0.66 \\
     \hline\label{table:1}
         \end{tabular}
  \end{center}
  \end{table}
  Figure \ref{fig:ent-kerr}  illustrates the behavior of $E_N$ when  the anharmonicity parameter, $\chi$, is varied.  It shows that for the cavity with only the  gain nonlinearity the amount of achievable optomechanical entanglement at the steady state is seriously limited by the stability condition. The Kerr nonlinearity makes the stability region larger, allows overcoming this limitation. 
 As before,  the reduction of the  entanglement    for $\chi=0.05$ (in comparison  to $\chi=0.03$) is related to the increment  of the   stability parameter $\eta_1$.
  
\begin{figure}[ht]
\centering
\includegraphics[width=3 in]{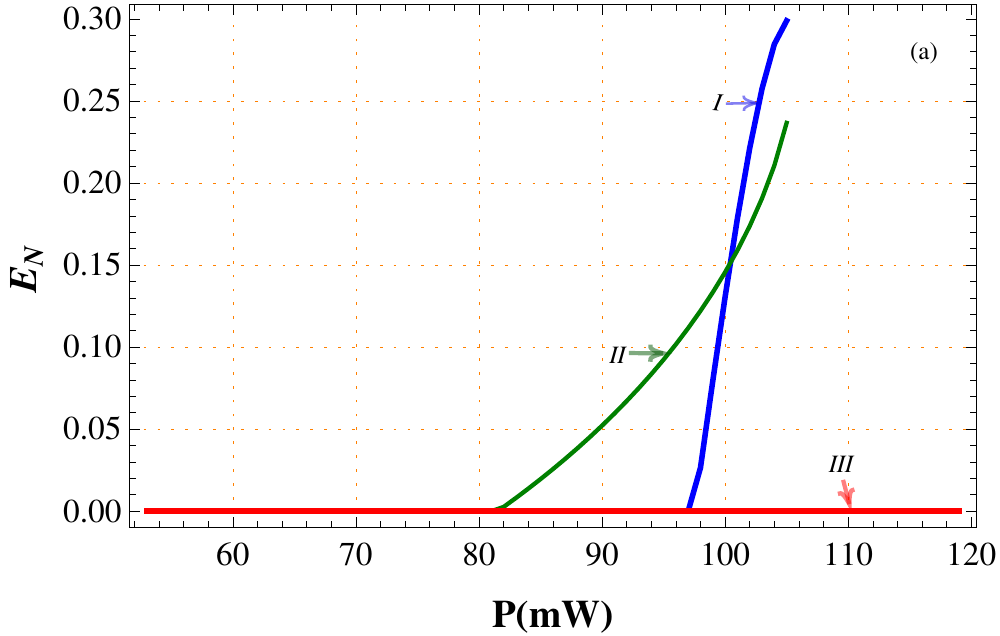}
\includegraphics[width=3 in]{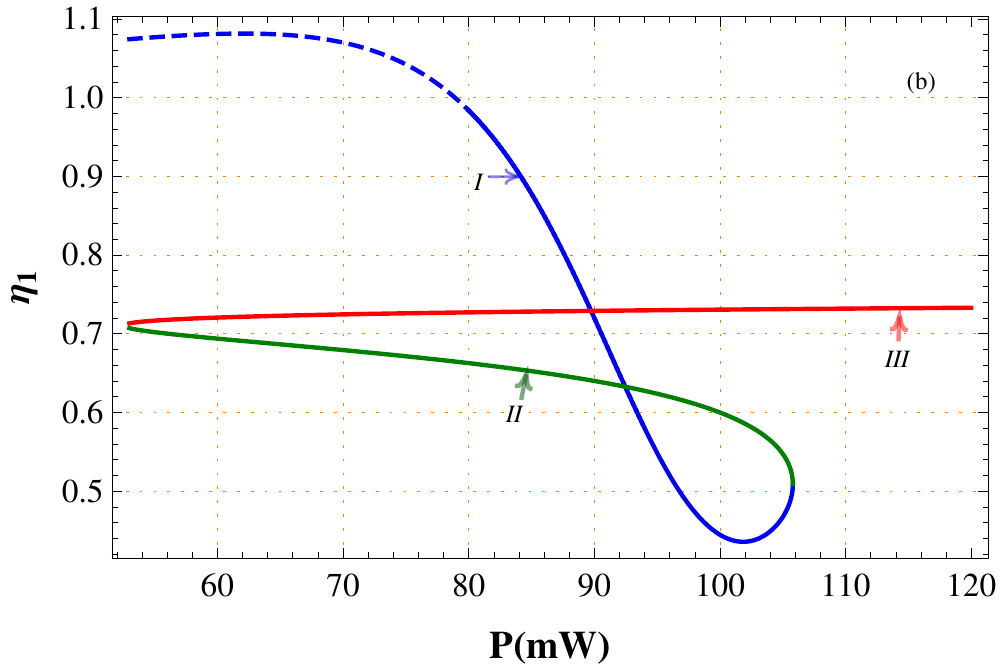}
\caption{(Color online) (a) The logarithmic negativity and (b) the stability parameter $\eta_1$ as  functions of the input power $P$  corresponding to the  three stable branches for $\theta=0.57\pi$ in Fig. \ref{fig:tristability}.
The dashed line corresponds to the unstable region.}
\label{fig:ent-tri}
\end{figure}

Until now we have studied the entanglement in the good cavity limit ($\kappa<\omega_m$) and   have chosen the  values of the  input power  and  the detuning far from the  multistability region. It is also interesting to examine the  entanglement in the bad cavity limit  and in the tristable regime. Figure \ref{fig:ent-tri}shows the  entanglement  and the stability parameter $\eta_1$ as  functions of  the laser input power $P$ in the tristable regime for the data of Fig. \ref{fig:tristability}. It shows that for the first and second branches the entanglement is maximum at the end of the  branches, while  for the third branch the phonon-photon  entanglement is null. This result might be interpreted as arising from the  Kerr-induced   shift of the resonance frequency of the cavity ( for the third branch this shift is more than $1.7\omega_m$).  Also, it can be seen that at the end of each stable branch $\eta_1\neq0$ and  the entanglement is found only in the region with $\eta_1<0.65$.

It should be noted that in order to stay within the range of validity of the linearization approximation,  we have been assured that  the condition $<\delta a^{\dagger}\delta a>_{ss}\ll a_s^2$  is satisfied in all the results given above.


\section{ Conclusions}\label{sum}
In conclusion, we have studied  the interaction of a single-mode field with both a weak Kerr medium and a parametric nonlinearity in an intrinsically nonlinear optomechanical system. We have investigated the stability  behavior of the intracavity mean photon number, the intensity, cooling and stationary entanglement.
  
We have found  that  the combination of the nonlinearities leads to a shift of  the  resonance frequency of the cavity toward the lower values and appearance of  three real roots for the steady-state response of the system. We have derived the general condition for tristability in the system.
Furthermore, we have studied the cooling and stationary entanglement  by using the covariance matrix formalism. In particular, we have shown that  the lower bound of the resolved sideband regime and the minimum attainable phonon number can be less than that of a bare cavity by controlling the parametric nonlinearity and the  phase of the field driving the OPA. The weak Kerr nonlinearity can be used to extend the domain of the stability to the desired range of the effective detuning in which the effective temperature of the system is minimized. Also, the  Kerr nonlinearity  modifies the behavior  of the stability parameter and  allows to approach significant entanglement simultaneously with the ground state cooling of the mirror. In the investigation of the degree of stationary  entanglement between the cavity and the mechanical modes we have identified  four key parameters  that affect the behavior of the entanglement. They are effective detuning of the cavity, the optomechanical coupling and the two stability parameters of the system. Also, as shown in this paper the present scheme allows to have significant entanglement in the tristable regime for the lower and middle branches which makes the current scheme distinct from the bare optomechanical system.

 \section*{Acknowledgements}
 S.Sh. is grateful to M. Xiao and R.Ghobadi for useful discussions.
The authors wish to thank The Office of Graduate Studies of The University of Isfahan for their support.
\bibliographystyle{apsrev4-1}

\end{document}